\documentclass{aastex631}

\usepackage{tikz}
\usepackage{hyperref}
\usepackage{float}
\renewcommand{\arraystretch}{1.5}

\begin{document}

\title{The weird and the wonderful in our Solar System:\\ Searching for serendipity in the Legacy Survey of Space and Time}
\author[0009-0008-2649-6426]{Brian Rogers}
\affiliation{University of Oxford, Department of Physics, Denys Wilkinson Building, Keble Road, Oxford, OX1 3RH, UK}
\affiliation{Astrophysics Research Centre, School of Mathematics and Physics, Queen's University Belfast, Belfast BT7 1NN, UK}
\author[0000-0001-5578-359X]{Chris J. Lintott}
\affiliation{University of Oxford, Department of Physics, Denys Wilkinson Building, Keble Road, Oxford, OX1 3RH, UK}
\author[0000-0003-4823-129X]{Steve Croft}
\affiliation{Breakthrough Listen, University of California, Berkeley, 501 Campbell Hall 3411, Berkeley, CA 94720, USA}
\affiliation{SETI Institute, 339 Bernardo Ave, Suite 200, Mountain View, CA 94043, USA}
\affiliation{University of Oxford, Department of Physics, Denys Wilkinson Building, Keble Road, Oxford, OX1 3RH, UK}
\author[0000-0003-4365-1455]{Megan E. Schwamb}
\affiliation{Astrophysics Research Centre, School of Mathematics and Physics, Queen's University Belfast, Belfast BT7 1NN, UK}
\author[0000-0002-0637-835X]{James R. A. Davenport}
\affiliation{Department of Astronomy, University of Washington, Seattle, WA 98195, USA}

\begin{abstract}
We present a novel method for anomaly detection in Solar System object data, in preparation for the Legacy Survey of Space and Time. We train a deep autoencoder for  anomaly detection and use the learned latent space to search for other interesting objects. We demonstrate the efficacy of the autoencoder approach by finding interesting examples, such as interstellar objects, and show that using the autoencoder, further examples of interesting classes can be found. We also investigate the limits of classic unsupervised approaches to anomaly detection through the generation of synthetic anomalies and evaluate the feasibility of using a supervised learning approach. Future work should consider expanding the feature space to increase the variety of anomalies that can be uncovered during the survey using an autoencoder.
\end{abstract}

\section{Introduction} 
\label{sec:intro}
Cataloguing the Solar System is a key science driver of the Legacy Survey of Space and Time \citep[LSST;][]{2019ApJ...873..111I}, which is to be conducted by the Vera C. Rubin Observatory. The scale of discovery will be astounding. With over 5 million new objects expected to be found during the survey \citep{2009EM&P..105..101J, 2018Icar..303..181J, 2009arXiv0912.0201L, 2010Icar..205..605S, 2015MNRAS.446.2059S, 2016AJ....151..172G, 2016AJ....152..103S, 2017AJ....154...13V, 2018arXiv180201783S, 2019ApJ...873..111I, 2020Icar..33813517F, Schwamb_2023}, we will soon have the largest catalog of Solar System objects that has ever existed. Outlined in the objectives of the Solar System Science Collaboration (SSSC) roadmap \citep{2018arXiv180201783S}, this catalog will be incredibly valuable for many areas of Solar System science. Many of the questions that LSST will answer have not yet been asked \citep{Schwamb_2023}. We will potentially find objects of new types and behaviours that will in turn, prompt new questions.

Previous work has considered the potential for discovering the unknowns in LSST time domain data \citep{Li_2022}. However, work is required to understand how anomaly detection may be performed effectively on Solar System data. For example, LSST has the potential to discover tens of interstellar objects (ISOs) \citep{Hoover_2022}. Colour outliers could reveal new types of fragments from known bodies or new surface types. Objects in very unusual orbits such as retrograde asteroids, or objects that demonstrate changes in their orbits \citep{Farnocchia_2023, 2018Natur.559..223M, 2017AJ....153...80H}, which could aid in our understanding of active processes in small bodies, are also discoverable. The most interesting examples, however, are those we have not yet imagined. Finding such objects in the large volume of data the survey will produce necessitates the development of effective anomaly detection systems; this paper demonstrates the efficiency of such a system using a simulation of the LSST survey output. 

Anomaly detection is the process of discovering patterns in data that do not conform to expected behaviour \citep{anom_detect}. Depending on the degree of information available about anomalies in a given data set, there are three main approaches to anomaly detection: supervised, semi-supervised, and unsupervised. With supervised detection, ideally each anomaly is identified and labelled. However, given the workload required in labelling each example, anomalies can be missed. The available instances can be used to train machine learning models that can detect anomalies, in our case unusual objects, not previously seen in training data. A large limitation of supervised detection approaches is that they can completely miss classes of anomalies that are not present in the training data.  In contrast, semi-supervised approaches instead assume that data used to train a model consists of non-anomalous instances. Unsupervised techniques require neither labels nor the assumption of normality in the dataset. Instead, they assume that normal instances are far more frequent than anomalies in the data. Methods then attempt to find the instances that are much different from the bulk of the data. We primarily consider unsupervised approaches in this work because they do not require the same training resources as supervised and semi-supervised methods, and so are more widely applicable to anomaly detection tasks. We briefly introduce a supervised approach in section \ref{section:deficiencies_unsupervised}.

We first consider a number of classic approaches to anomaly detection. Using a subspace of possible object features including orbital and colour properties, in section \ref{ref:classic_approaches_to_anomaly_detection} we generate synthetic anomalies to evaluate a variety of unsupervised approaches. Then in section \ref{section:deficiencies_unsupervised}, we present how supervised anomaly detection can improve on the limitations of the classic unsupervised methods, even with modest training data.

We then improve on these approaches using deep learning. Section \ref{section:deep_ae} demonstrates how a deep autoencoder can be applied to object anomaly detection. Introduced by \citet{2006Sci...313..504H}, deep autoencoders can learn a compressed representation of the feature space, effectively reducing the dimensionality of the data. Deep methods for dimensionality reduction and anomaly detection have been considered before for astronomical applications. \citet{VAE_SDSS} considered the application of vanilla but primarily variational autoencoders to spectra obtained by the Sloan Digital Sky Survey. However, we are unaware of any previous approaches using deep anomaly detection in the Solar System domain.

We adopt the vanilla autoencoder for this work because we are interested primarily in the reconstruction loss, unlike in other autoencoder variations which consider additional factors. In section \ref{section:creating_ae}, we reduce the dimensionality of the feature space to create a meaningful latent space. We explore two key aspects of the autoencoder approach: firstly, the reconstruction loss applied to anomaly detection, in section \ref{section:recon_loss}; and secondly, similarity searching to look for additional interesting objects, in section \ref{section:sim_search}.

 With the diversity of Solar System objects expected in the LSST survey, what constitutes a interesting anomaly will inevitably vary from use case to use case, and from user to user. We therefore must design a system which performs well regardless of the type of anomaly we are interested in. Constructing an agnostic anomaly detection system for objects given the number of their features to consider is a challenging task. We explore how the autoencoder can be used as a foundation for creating more complex and personalised detection systems in section \ref{section:ae_ensembling}, before commenting on future uses of these technologies.

\section{Data}
\label{section:}

This paper makes use of two simulations of the LSST solar system data expected at the end of the full, ten-year, survey. These are generated using rubin\underline{\hspace{.1in}}sim cadence simulations \citep{2022zndo...7374619Y}. Both simulations are based on synthetic objects generated using the Synthetic Solar System Model \citep[S3M;][]{S3M}, in addition to already known objects. The simulations account for models of cadence, sensitivity and completeness, and though they are works in progress are sufficient here for assessing the performance of anomaly detection algorithms.

In section \ref{ref:classic_approaches_to_anomaly_detection}, we use the Solar System Products Database accessed through the LINCC (LSST Interdisciplinary Network for Collaboration and Computing) Jupyter hub. The simulation, which utilises \textit{v1.7} of the baseline cadence,  consists of four tables containing synthetic objects\footnote{ \url{https://github.com/dirac-institute/hybrid_sso_catalogue}}: The SSObject table contains information for each object discovered in the survey, which is supplemented by the MPCORB table \citep{wagg} that includes all objects known by the Minor Planet Center (MPC)\footnote{\url{https://www.minorplanetcenter.net/}}. The SSSource and DiaSource contain information for each detection event for these objects. The information contained in each of these tables is listed in the LSST data products definition document \citep[DPDD; ][]{LSE-163}.

In section \ref{section:deep_ae}, we use the Rubin Data Preview 0.3 (DP0.3) version of the simulation released in August 2023 through the Rubin Science Platform (RSP) which was generated using \textit{v3.0} of the baseline cadence. The data were generated by  members of the Rubin Solar System Processing team, with help from the SSSC. This version has the same schema as the Solar System Products Database. However, it is composed of two sets of catalogs containing real and simulated Solar System and interstellar objects. These catalogs are combined and form the full 10 year survey simulation, which we use in this paper.

DP0.3 contains objects belonging to a variety of small body populations, including: near Earth objects, main belt asteroids, Trojans, Hildas, trans-Neptunian objects, interstellar objects and long period comets. The simulated survey detects 3.2 million S3M objects and 2,429 simulated ISOs. Further it detects 3,344 long-period comets, which were generated using a model developed by \citet{2019AJ....157..181V}\footnote{The full list of the long period comet populations are available from the following: \url{https://github.com/lsst-sssc/SSSC_test_populations_gitlfs}}. These objects are simulated using two colour classes: silicaceous (S) and carbonaceous (C). The catalog of synthetic objects is combined with a real object population to generate a realistic version of the catalog expected from the survey\footnote{\url{https://dp0-3.lsst.io/data-products-dp0-3/data-simulation-dp0-3.html}}. It is important to note that DP0.3 does not necessarily represent the most likely output of the LSST Solar System survey; it contains 20$\%$to 30$\%$ fewer objects than expected, for example. In addition, the number of ISO and long period comets exceeds predictions, as these populations were inflated on purpose to allow for a diverse range of properties to be included. Much more work is underway to prepare realistic, well understood survey populations.

However, for our purposes all we require is a simulation which reflects the diversity of objects in the final survey; techniques which perform well on this simulation can be tested on more realistic outputs once they become available. In the meantime, this work does not attempt to quantify the likely yield of unusual objects, but merely to demonstrate that we can find them in a large survey of the type which will be produced by LSST.
 
\section{Classic approaches to anomaly detection.}
\label{ref:classic_approaches_to_anomaly_detection}
There are a wide range of approaches to anomaly detection. We evaluate the performance of a series of popular unsupervised anomaly detection methods using synthetically generated anomalies, of which we will consider three general types: global, cluster, and local. We then consider the limitations of these methods, comparing to a supervised anomaly detection approach.

\subsection{Methods}

\subsubsection{Orbit-colour feature space}
325,000 Solar System objects with semi-major axis less than 6\,au were randomly selected from the DiRAC simulation using the SSObject and MPCORB tables. Initially we restrict the objects we consider to the inner Solar Solar system. We consider five features from the tables which collectively make up the orbit-colour space. This feature space includes: semi-major axis (a), orbital eccentricity (e), the sine of the orbital inclination ($sin (i)$), the i-z colour, and the $a^*$ composite colour. The $i - z$ colour is the difference between the fitted $i$-band and $z$-band magnitudes for the object. The $a^*$ composite colour is calculated as 
$$ a^* = 0.89(g-r) + 0.45(r-i) - 0.57$$
where $g - r$ is the difference between the $g$ and $r$-band absolute magnitudes and $r - i$ is the difference between the $r$ and $i$-band magnitudes. This is an optimal asteroid colour developed by \citet{2001AJ....122.2749I} using principal component analysis. Previous studies \citep{2002AJ....124.2943I, Parker_2008} have also considered the $a^*$ and $i - z$ colours for the analysis of asteroid families. They both made use of a similar parameter space but instead used the proper orbital elements, whereas we use the osculating orbital elements provided by the MPC. Given the usefulness of this subspace for previous studies of small bodies, we adopt it in this section for application to anomaly detection.

\subsubsection{Gaussian mixture modelling of the orbit-colour space.}
\label{section:GMM}
\begin{figure}
    \centering    \includegraphics[width=\textwidth]{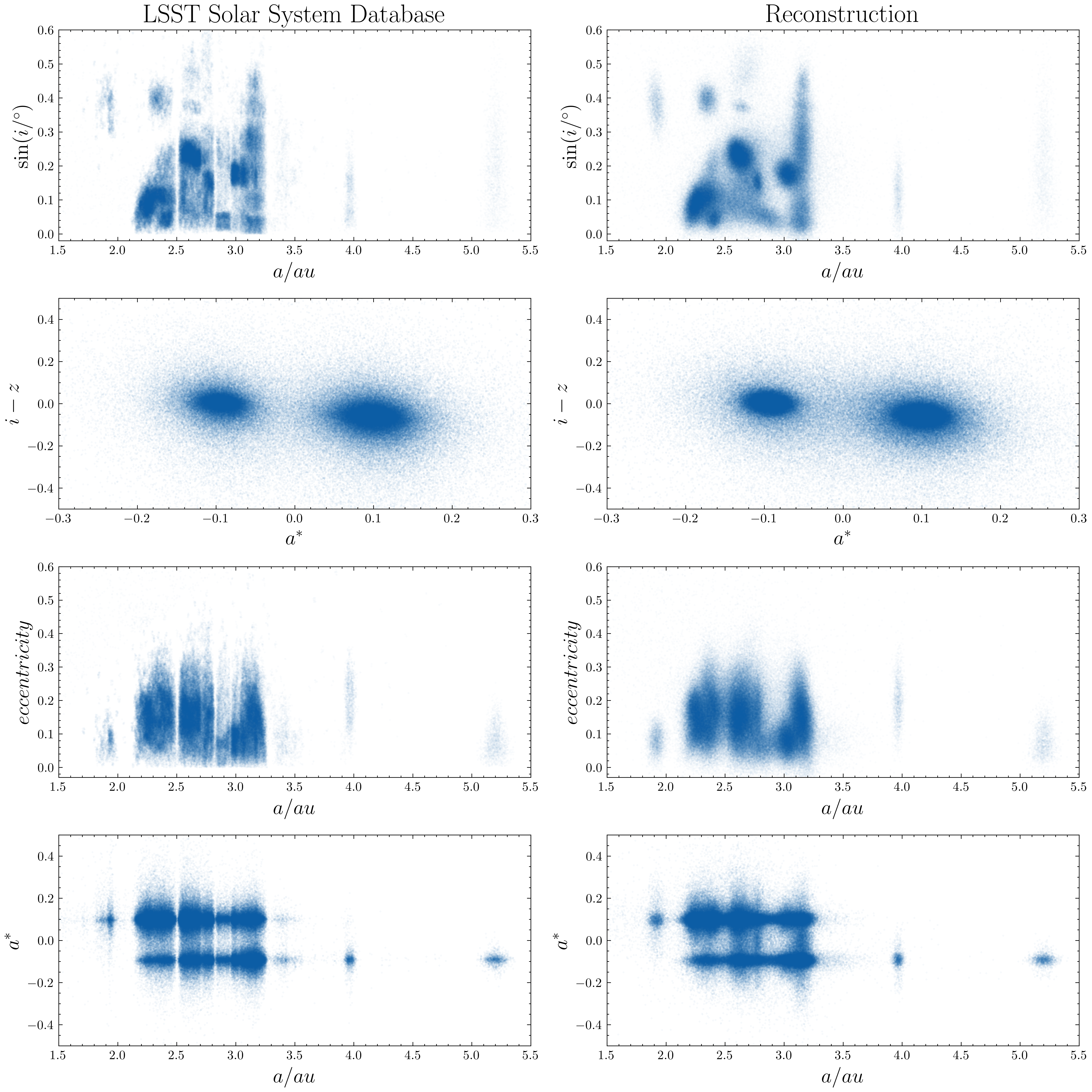}
    \caption{Reconstruction of the orbit-colour space using the Gaussian mixture model.}
    \label{fig:gmm_recon}
\end{figure}

In order to assess the performance of unsupervised anomaly detection techniques, a test data set of anomalies is required. We trained a Gaussian mixture model (GMM) on the orbit-colour space using the Sci-Kit learn package \citep{scikit-learn}. 
The GMM is an effective way to model the feature space by using a finite mixture of multivariate Gaussian distributions. By using a sufficient number of Gaussians, complex density functions can be approximated to arbitrary accuracy \citep{10.5555/1162264}. Figure \ref{fig:gmm_recon} shows how this learned GMM can reconstruct the feature space. Each Gaussian distribution in the mixture model, also called a component, has a mean and covariance. The Sci-Kit learn implementation of the mixture model uses the expectation-maximization (EM) algorithm to find the optimal means and covariances for each component in the model. The expectation-maximization is a general technique for finding maximum likelihood solutions for probabilistic models that have latent variables \citep{10.5555/1162264}.

We used a grid search to find the optimal number of components for the mixture model. We used the Bayesian information criterion \citep[BIC;][]{BIC} to perform this. By adding more components to the mixture model, it is possible to increase the maximum likelihood of the model. However, this can quickly result in overfitting. The BIC seeks to add a penalty term for the number of components used in the model in order to prevent the problem of overfitting. The optimal number of mixture components was determined to be 71. There is some physical intuition for a large number of components being required. Previous studies of the main belt population have revealed that many asteroids are related in families that have common origins. \citet{1918AJ.....31..185H} originally presented this idea by using orbital elements, and \citet{2002AJ....124.2943I} and \citet{2008Icar..198..138P} extended this analysis to use the $a^*$ and $i - z$ colours\footnote{See also recent work by \citet{2022A&A...664A..51R}}. As these families of asteroids have similar orbital elements and colours, they cluster together in the orbit-colour space. This lends support to the idea that a large number of our model components correspond to particular asteroid families. However, we cannot perform cross correlation of known objects that belong to families, because objects in this version of the simulation cannot be directly linked to real objects. Further, the colour distribution used in the two simulations is simplified (constrained to Gaussian distributions centered around the average C-type and S-type colors) compared to the real distribution.

The learned GMM can then be used as a generative model by sampling from the mixture components. We manipulated these components to generate anomalous samples that could be used to benchmark the performance of anomaly detection methods. This is an approach previously used by \citet{han2022adbench} for evaluating anomaly detection methods on a series of 57 benchmark anomaly datasets across various practical application domains. This procedure ensures we can make systematic choices for the given anomaly detection task, in order to find the optimal method. There are other important benefits to this approach. Complex anomaly types can be generated without making assumptions about their underlying distribution. Additionally, as the GMM is a probabilistic model, the log-likelihood score can be used to quantify how likely a given object is to be generated by the model. This means that any object, whether in the data or generated synthetically, can be ranked and scored according to its outlier properties.

\label{section:synthetic_anom_gen}

Global anomalies lie outside of the distribution of normal objects such that they appear to originate from a different distribution. In order to simulate this anomaly type, we draw from a uniform distribution, $U$, given by

\begin{equation}
    X_{i} = U(\mu_{i} - \alpha \cdot \sigma_{i}, \mu_{i} + \alpha \cdot\sigma_{i}),
\end{equation}

where $X_{i}$ is the feature vector of an object, $\mu_{i}$ and $\sigma_{i}$ are the mean and standard deviation of the $i^{th}$ feature, and $\alpha$ is a scaling factor that controls how anomalous the generated anomalies are. We choose $\alpha = 3$ to provide a conservative limit on the global anomalies we aim to detect. An asteroid that lies significantly outside the orbit-colour properties of any nearby objects would be considered a global anomaly.

Cluster anomalies are generated by selecting mixture components at random and multiplying their mean by a scale factor $\beta$. We chose $\beta = 1.2$ which corresponds to a maximum 20$\%$ shift in any of the means of a mixture component from its original centroid. We would consider an emerging new family as a cluster anomaly in the orbit-colour space.

Local anomalies are more subtle than the preceding two anomalies and therefore harder to detect. They are generated by scaling the covariances of the mixture model by a scalar factor $\gamma$. We chose $\gamma = 16$ so that the standard deviation, $\sigma_i$ of the $i^{th}$ feature is scaled by a factor of 4 compared to the original distribution. This has the effect of widening and populating the edges of the distribution between $3\sigma_i$ and $4\sigma_i$, a key feature of local anomalies. By scaling by much higher factors, local anomalies begin to replicate the behaviour of global anomalies. By using lower scaling factors, the generated samples are insufficiently different from the original distribution. An unusual member within an asteroid family is one practical example of a local type anomaly. 

\begin{figure}
    \centering
\includegraphics[width=\textwidth]{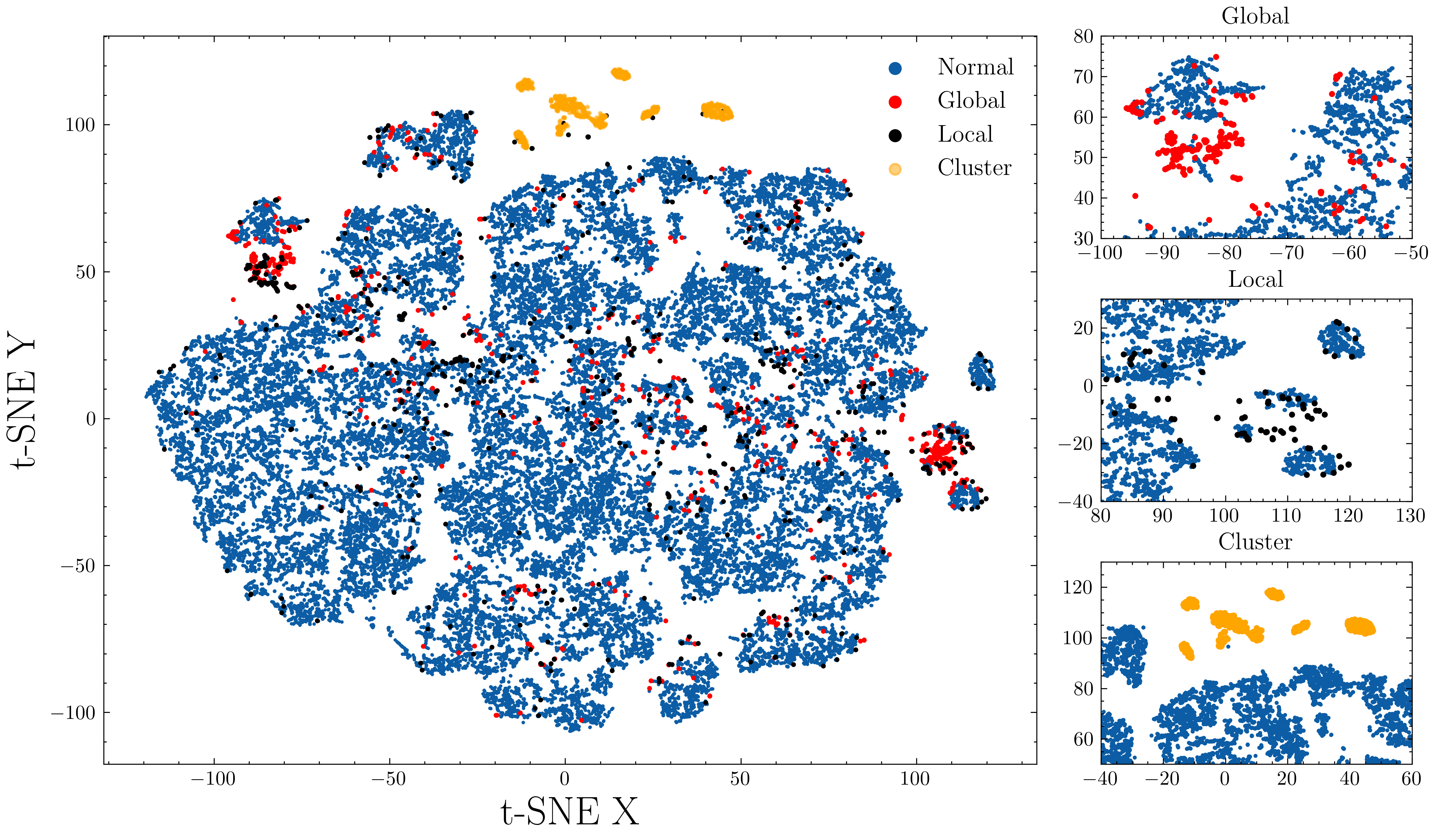}
    \caption{The generated anomalies visualised using t-SNE. Global anomalies (red) are interwoven between the normal objects (blue). The clustered anomalies (orange) lie in dense groups around the edges of the normal objects in this reduced space. Local anomalies (black) tend to group round the edges of normal clusters of points.}
    \label{fig:anom_visual}
\end{figure}

A graphical representation of the anomalous points generated can be seen in the 2D projection (Figure \ref{fig:anom_visual}) of the space using T-distributed Stochastic Neighbor Embedding \citep[t-SNE;][]{tsne}. t-SNE is a popular dimensionality reduction technique which is particularly useful for visualising high dimensional data. In our case it is useful for visualising the generated synthetic anomalies. Global anomalies (red) lie in less dense regions of the parameter space whilst local anomalies (black) form around the edges of clusters of normal points (blue). Cluster anomalies (orange) can be seen scattered around the edges of the core of normal points.

Each anomaly was labelled using its type for classification. With the Gaussian mixture model of the orbit-colour space, we can also quantify of how much of an outlier an object is using the log-likelihood score. Comparisons can then be performed on the scores produced by the outlier detection algorithms, in order to evaluate which method yields the greatest number of interesting anomalies.

Using the log-likelihood scores, a threshold was applied to the normal training data to ensure no unlabelled anomalous samples were in the training set. We defined normal objects as falling within $3 \sigma$ from the mean log-likelihood score, removing any normal points that broke this threshold, which we then generated 50,000 normal objects that met this criteria. This allows local anomalies to be generated and populated beyond 3$\sigma$ in accordance with their definition. 1000 anomalies in each class were generated. Anomalous objects with a semi major axis that exceeded than 6 \,au were removed as we only consider this particular inner region of the Solar System. Furthermore, generated anomalies that did not have values of $-1 \le sin (i) \le 1$ were also removed. Using these criteria, local anomalies accounted for 1.5\% of the normal object and local anomaly combined dataset. Further, we ensured that the generated anomalies were truly anomalous by setting a maximum threshold of 0 on the log-likelihood scores of the global and cluster anomalies. This implies that any generated global and cluster anomalies are at least 4$\sigma$ from the mean log-likelihood score. After generating these objects from the above description, global anomalies accounted for 0.8\% of the normal and global combined dataset whereas cluster anomalies accounted for 1.9\% of the combined normal and cluster dataset.

Using the outlier detection library PyOD \citep{zhao2019pyod}, which provides efficient implementations of a series of unsupervised anomaly detection methods, we found how each method performed on a given anomaly type. We used the default hyper-parameters from PyOD to ensure a fair baseline comparison between methods. We used nine methods from the PyOD library: Angle based outlier detection (ABOD), clustering based local outlier factor (CBLOF), feature bagging, Cook's distance (CD), Copula-based outlier detection (COPOD), histogram-based outlier score (HBOS), isolation forest (IForest),  k-nearest neighbors (KNN), and local outlier factor (LOF). Literature references for these methods, and the results from running them on our dataset, are presented in Section~\ref{section:unsupervised_detection_results}.

Each method was evaluated on balanced accuracy, F1 score, and ranking. For the binary classification of anomalies there are four possible output types from a detection model:
\begin{itemize}
    \item True positive (TP) - the model correctly predicts this is an anomalous instance.
    \item True negative (TN) - the model correctly predicts this is not an anomalous instance.
    \item False negative (FN) - the model incorrectly predicts that the instance is not an anomaly.
    \item False positive (FP) - the model incorrectly predicts that the instance is an anomaly.
\end{itemize}
From these the balanced accuracy can be written as
\begin{equation}
    Accuracy = \frac{1}{2} \left[ \frac{TN}{TN+FP}  + \frac{TP}{TP+FN} \right]
\end{equation}

The F1 metric, which is the harmonic mean of the precision and recall of the model, can be calculated by 
\begin{equation}
    F1 = \frac{2*Precision*Recall}{Precision+Recall} = \frac{2*TP}{2*TP+FP+FN}
\end{equation}

The ranking metric calculated what ratio of the highest 500 scored anomalous instances were in the top 500 instances suggested by the given method's score. For global and cluster anomalies, the log-likelihood can be used directly as an outlier score since we are drawing from a newly created distribution. However, this methodology cannot be transferred to local anomalies because they are sampled from the model's components. We instead rank local anomalies by the Euclidean distance from the centroid of the component they belong to.

\subsection{Unsupervised detection results}
\label{section:unsupervised_detection_results}
The results for each anomaly type are presented in Tables~\ref{table:Global}, \ref{table:Cluster}, and \ref{table:local}. For global anomalies the k-Nearest Neighbors (KNN) gave the strongest performance in accuracy, F1 and ranking metrics. However, when applied to clustered anomalies its performance degraded to no better than random selection. Here we see the importance of choosing the `right tool for the job' when searching for anomalies. 
\begin{deluxetable}{lccc}[H]
\label{table:Global}
\tablehead{
\colhead{Algorithm} & \multicolumn{3}{c}{Global} \\ \colhead{} & \colhead{Accuracy} & \colhead{F1} & \colhead{Ranking}
}
\startdata
Angle Based Outlier Detection \\ \citep[ABOD; ][]{ABOD} & 0.718 $\pm$ 0.032 & 0.443 $\pm$ 0.044 & 0.723 $\pm$ 0.009\\
Clustering-Based Local Outlier Factor \\ \citep[CBLOF; ][]{CBLOF} & 0.504 $\pm$ 0.005 &  0.015 $\pm$ 0.001 & 0.397 $\pm$ 0.033\\
Cook's Distance \\ \citep[CD; ][]{CD} & 0.648 $\pm$ 0.027 &  0.268 $\pm$ 0.025 & 0.541 $\pm$ 0.015 \\
Copula-Based Outlier Detection \\ \citep[COPOD; ][]{COPOD} & 0.645 $\pm$ 0.016 &  0.293 $\pm$ 0.031 & 0.478 $\pm$ 0.018 \\
Feature Bagging \\ \citep{FeatureBagging} & 0.680 $\pm$ 0.030 & 0.363 $\pm$ 0.059 & 0.484 $\pm$ 0.028\\
Histogram-based Outlier Score \\ \citep[HBOS; ][]{HBOS} &  0.719 $\pm$ 0.022 &  0.442 $\pm$ 0.029 & 0.518 $\pm$ 0.022\\
Isolation Forest \\ \citep[IForest; ][]{IForest} & 0.684 $\pm$ 0.025 & 	0.368 $\pm$ 0.040 & 0.498 $\pm$ 0.033\\
k-Nearest Neighbors \\ \citep[KNN; ][]{KNN} & 0.812 $\pm$ 0.020 & 0.623 $\pm$ 0.025 & 0.770 $\pm$ 0.010\\
Local outlier factor \\ \citep[LOF; ][]{LOF} & 0.690 $\pm$ 0.017 &  0.382 $\pm$ 0.0257 & 0.419 $\pm$ 0.021\\
\enddata
\caption{The performance results of anomaly detection models using synthetically generated global anomalies.}
\end{deluxetable}

\begin{deluxetable}{lccc}[H]
\label{table:Cluster}
\tablehead{
\colhead{Algorithm} & \multicolumn{3}{c}{Cluster} \\ 
\colhead{} & \colhead{Accuracy} & \colhead{F1} & \colhead{Ranking}
}
\startdata
ABOD & 0.491 $\pm$ 0.001 & 0.001 $\pm$ 0.002 & 0.341 $\pm$ 0.006 \\
CBLOF & 0.855 $\pm$ 0.0147 &  0.718 $\pm$ 0.019 & 0.659 $\pm$ 0.012\\
CD & 0.871 $\pm$ 0.024 &  0.715 $\pm$ 0.032 & 0.683 $\pm$ 0.018\\
COPOD & 0.678 $\pm$ 0.008 & 0.368 $\pm$ 0.0126 & 0.483 $\pm$ 0.016\\
Feature Bagging & 0.519 $\pm$ 0.012 & 0.056 $\pm$ 0.024 & 0.298 $\pm$ 0.020\\
HBOS & 0.701 $\pm$ 0.023 &  0.412 $\pm$ 0.039 & 0.487 $\pm$ 0.026\\
IForest & 0.890 $\pm$ 0.042 &  0.784 $\pm$ 0.089 & 0.684 $\pm$ 0.014\\
KNN & 0.490 $\pm$ 0.000 &  0.000 $\pm$ 0.000 & 0.363 $\pm$ 0.012\\
LOF & 0.499 $\pm$ 0.004 &  0.017 $\pm$ 0.007 & 0.216 $\pm$ 0.013
\enddata
\caption{The performance results of anomaly detection models using synthetically generated cluster anomalies.}
\end{deluxetable}
Considering cluster anomalies, Isolation Forest (IForest) performed best across each metric with Cook's distance (CD) also performing relatively strongly. 

\begin{deluxetable}{lccc}[H]
\label{table:local}
\tablehead{
\colhead{Algorithm} & \multicolumn{3}{c}{Local} \\ 
\colhead{} & \colhead{Accuracy} & \colhead{F1} & \colhead{Ranking}
}
\startdata
ABOD & 0.723 $\pm$ 0.021 & 0.460 $\pm$ 0.030 & 0.288 $\pm$ 0.013  \\
CBLOF & 0.578 $\pm$ 0.008 &  0.171 $\pm$ 0.016 & 0.625 $\pm$ 0.036 \\
CD & 0.739 $\pm$ 0.037 &  0.252 $\pm$ 0.082 & 0.460 $\pm$ 0.057 \\
COPOD & 0.648 $\pm$ 0.014 &  0.305 $\pm$ 0.020 & 0.379 $\pm$ 0.012 \\
Feature Bagging & 0.733 $\pm$ 0.021 & 0.464 $\pm$ 0.021 & 0.244 $\pm$ 0.033 \\
HBOS &  0.652 $\pm$ 0.023 &  0.310 $\pm$ 0.035 & 0.278 $\pm$ 0.025 \\
IForest & 0.654 $\pm$ 0.016 & 0.314 $\pm$ 0.028 & 0.477 $\pm$ 0.025\\
KNN & 0.730 $\pm$ 0.014 & 0.544 $\pm$ 0.018 & 0.306 $\pm$ 0.015 \\
LOF & 0.720 $\pm$ 0.015 &  0.445 $\pm$ 0.020 & 0.203 $\pm$ 0.012\\
\enddata
\caption{The performance results of anomaly detection models using synthetically generated local anomalies.}

\end{deluxetable}

Searching for local anomalies had a much more mixed performance. A number of methods had similar performance metrics including k-nearest neighbor (KNN), local outlier factor (LOF), feature bagging, Cook's distance (CD) and angle based outlier detection (ABOD). This indicates that performance on local anomaly classification is largely invariant to the choice of these methods. However for the ranking metric, which we defined as the distance of the instance to its component's centroid, the clustering based local outlier factor (CBLOF) had the best performance. Yet it performs poorly on the classification task. This highlights that the choice of detection tool is dependent on how we rank what anomalies are interesting or not. This is a problem when trying to construct an agnostic anomaly detection system as one method will not suit all possible tasks nor all user definitions of what constitutes an anomalous instance.

\subsection{The deficiencies of standalone unsupervised methods.}
\label{section:deficiencies_unsupervised}

In addition to the problem of defining what exactly is an interesting anomaly, the evaluation tables (\ref{table:Global}, \ref{table:Cluster}, and \ref{table:local}) make it clear that the even the best classic unsupervised methods we considered can perform poorly on the F1 metric and the balanced accuracy. This means that many false positive anomaly examples may be flagged wasting a user's valuable time, especially in the case of low balanced accuracy scores which has a much greater effect on the false positive rate. Worse still, a high rate of false negatives means that anomalies can be missed. This problem is made more acute by examining the ranking metrics, as some of the most interesting anomalies are missed by the unsupervised methods. Missing anomalous examples, especially the most interesting cases, is intolerable for any efficacious anomaly detection system.

Supervised detection methods promise much better performance. However they require labelled training data which necessitates human input. In order to evaluate whether it is feasible to apply supervised techniques to this anomaly detection task, we examined how many positive anomaly training labels a classifier would require to outperform the best unsupervised methods on global anomalies.

We used XGBoost \citep{XGBoost}, a popular boosting method to perform binary classification: simply, is this object an anomaly or not? Examples of global anomalies were generated, selected randomly and injected into a set of normal objects. The number of anomalies used to train the classifier was then varied.

\begin{figure}
    \label{fig:active_learing}
    \centering
\includegraphics[width=\textwidth]{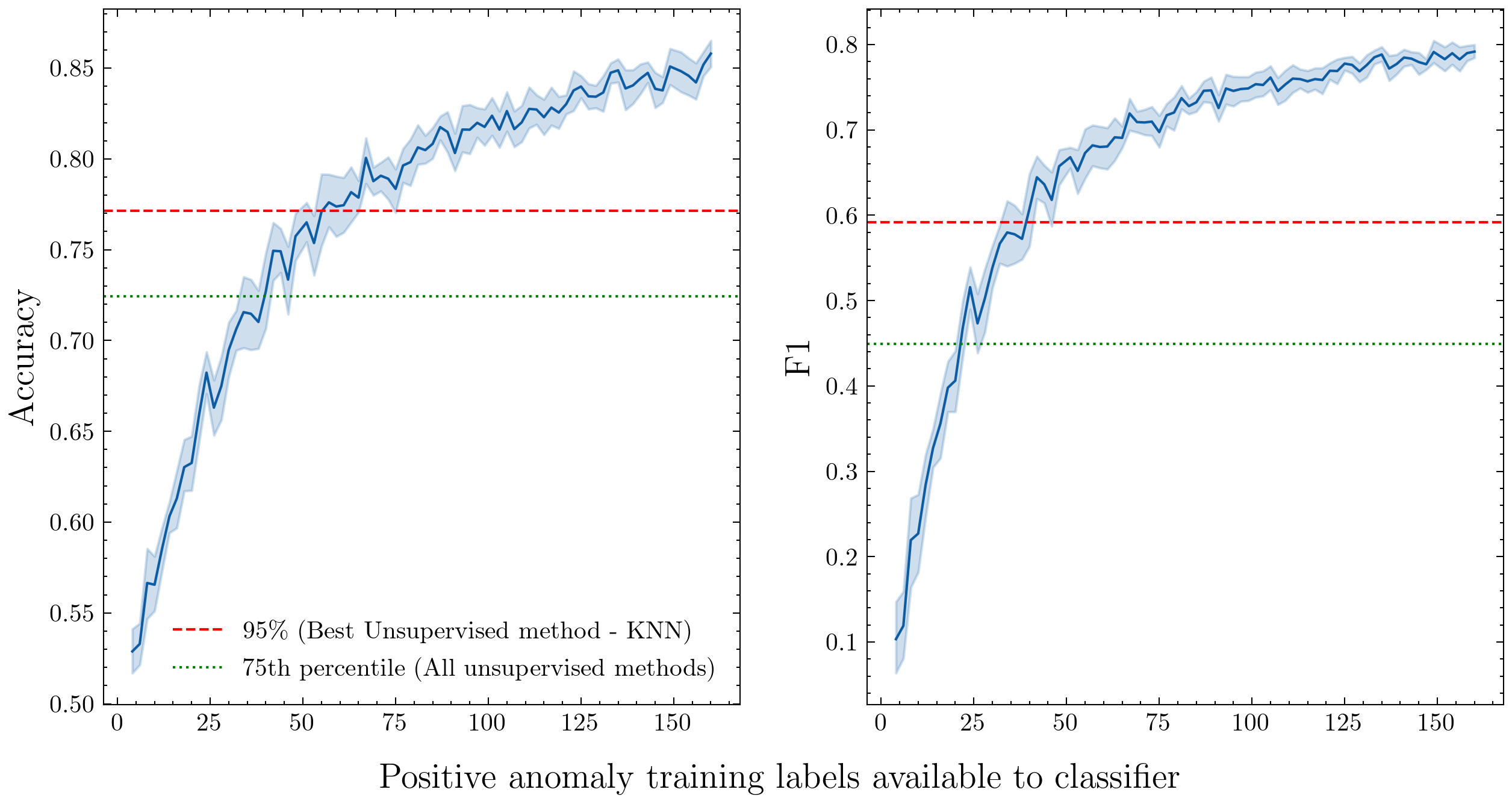}
    \caption{The accuracy and F1 metrics for the XGBoost classifier for given a number of training labels. The horizontal red line is 95$\%$ of the score of the best performing unsupervised method (KNN) on global anomalies. The green line is the 75th percentile performance of all unsupervised detection methods considered.}
    \label{fig:active_learning}
\end{figure}

Figure \ref{fig:active_learing} shows the performance of the supervised classifier as a function of available labelled anomalies in the training set.  In order to achieve 95$\%$ of the accuracy and F1 scores of the best unsupervised method, the classifier required only 57 and 40 labelled examples respectively. The improvement of the F1 score is especially notable. This suggests that even modest levels of human input into the anomaly detection procedure can significantly improve output quality. This is important for two reasons. Firstly, decreasing the number of false negatives means fewer anomalous instances will be missed. Secondly, decreasing the number of false positives produced by the method saves valuable time for users who are looking for the most interesting objects. 

Anomalous examples must be found in the first place to train supervised models. Such examples are required for each type, substantially increasing the effort required. This can be achieved by using unsupervised methods to filter the bulk of the data and presenting the highest machine scoring anomalies to a user. The most interesting anomalies can then be selected by a user for training a supervised model. The performance of the unsupervised method will determine what anomalies should be presented to a user, therefore it is important that a highly effective unsupervised method is chosen, otherwise this can be a substantial limitation. From this dual approach, further examples can be found and the supervised model can be refined in an iterative loop of self-improvement, to help overcome the shortcomings of the initial unsupervised method. This approach, referred to as active learning, has been previously demonstrated by \citet{Lochner_2021} to been effective in increasing the number of interesting anomalies that are returned to a user. This approach is also particularly beneficial as these models can be personalised per user or use case.

This section has considered how effective anomaly detection can be performed on a particular subspace of possible object parameters, the orbit-colour space, using classical approaches. However, the variety of anomalies detected are limited to this selection of features. Considering more features and therefore detecting a greater variety of anomalies requires increased computational resources and encounters difficulties with computational efficiency.

Even with additional resources, the performance of these methods can suffer in high dimensional spaces. Noisy features can mask more relevant features. Selecting the right features for our chosen subspace is a problem which grows with the number of dimensions we include. Choosing to split these feature spaces into smaller subspaces, i.e. consider smaller parts of the parameter space for outliers, has associated problems. \cite{https://doi.org/10.1002/sam.11161} define a version of the `curse of dimensionality' such that given enough subspaces to consider, it is possible to find a subspace in which any object appears to be an outlier. This is problematic if we want to create a general detection system, capable of finding interesting objects that are anomalous in a meaningful sense to users of the data products.

Given that these methods will not scale to include the full feature space, we turn to a deep learning approach in section \ref{section:deep_ae} to address these problems. 

\section{Deep autoencoder approach to anomaly detection.}
\label{section:deep_ae}
In the following section, we explore how a deep autoencoder can be applied to anomaly detection on Solar System object data in the era of LSST. Autoencoders have a series of useful properties that address the problems encountered in Section~\ref{section:deficiencies_unsupervised}. Using an autoencoder, we can reduce the dimensionality of the feature space but retain meaningful properties of the LSST data. This is an important consideration given the volume and depth of the Solar System data expected from the survey and with the problems of high dimensional feature spaces highlighted in section~\ref{section:deficiencies_unsupervised}. Importantly for anomaly detection, the reconstruction loss associated with an autoencoder can be used directly for anomaly detection applications. Finally, the autoencoder provides a latent space representation that can be readily searched for objects. We explore these features in the following sections.

\subsection{Methods - creating the autoencoder.}
\label{section:creating_ae}
We select object features from the SSObject and MPCORB tables. Using each filter measurement, we compute colour indicies across the $griz$ filter bands. The u and y filter bands are ignored due to the low signal to noise ratio expected in these bands during the survey. We further augment the feature space by adding the semi-major axis\footnote{For objects with $e > 1$ (i.e.\ interstellar objects), we calculate a semi-major axis which is used as a feature in the autoencoder as was performed for objects with $e < 1$}. We emphasise this is not physical property of the body, rather a value must be imputed in order for the object to be input into the autoencoder for training and inference., aphelion, and Jupiter tisserand parameter. This latter parameter quantifies the effect of Jupiter on a body's orbit and it is useful for distinguishing between different orbit types, especially for Jupiter family comets. Including other orbital elements, magnitude and phase slope parameters and their respective error measurements, we consider a total of 35 features (table \ref{table:features}).

\begin{deluxetable}{lc}
\label{table:features}
\tablehead{
\colhead{Name} & \colhead{Description} 
}
\startdata
a & Semi-Major axis (au)\\
i & Inclination to the ecliptic, J2000.0 (degrees) \\
e & Eccentricity \\
q & Perihelion (au) \\
Q & Aphelion (au) \\
peri & Argument of perihelion, J2000.0 (degrees) \\
node & Longitude of the ascending node, J2000.0 (degrees) \\
Flags & Bitwise flags indicating failure in phase curve fits \\
H & Absolute magnitude in V-band \\
gH & g band best fit absolute magnitude\\
gHErr & Uncertainty in gH\\
gG12 & g band best fit $G_{12}$ slope parameter\\
gG12Err & Uncertainty in gG12 \\
gChi2 & $\chi^2$ statistic of the phase curve fit (g band) \\
rH & r band best fit absolute magnitude\\
rHErr & Uncertainty in rH\\
rG12 & r band best fit $G_{12}$ slope parameter\\
rG12Err & Uncertainty in rG12 \\
rChi2 & $\chi^2$ statistic of the phase curve fit (r band) \\
iH & i band best fit absolute magnitude\\
iHErr & Uncertainty in iH\\
iG12 & i band best fit $G_{12}$ slope parameter\\
iG12Err & Uncertainty in iG12 \\
iChi2 & $\chi^2$ statistic of the phase curve fit (i band) \\
iNdata & Number of data points used to fit the i band phase curve \\
zH & z band best fit absolute magnitude\\
zHErr & Uncertainty in zH\\
zG12 & z band best fit $G_{12}$ slope parameter\\
zG12Err & Uncertainty in zG12 \\
zChi2 & $\chi^2$ statistic of the phase curve fit (z band) \\
i - z & Difference between iH and zH fitted magnitudes\\
g - r & Difference between gH and rH fitted magnitudes\\
r - i &  Difference between rH and iH fitted magnitudes\\
a$^*$ &  $0.89(g-r) + 0.45(r-i) - 0.57$\\
\\
$T_{J}$ & Tisserand's  parameter (Jupiter)  \\
\\
& $\frac{a_{J}}{a} + 2\cos(i)\sqrt{\frac{a}{a_{J}}(1 - e^2)}$ \\
\\
& where $a_J$ is the semi-major axis of Jupiter\\
\enddata
\caption{Input features and their description. These features are taken from the SSObject and MPCORB table of the simulation.} See footnote \footnote{The iNdata has been used to inform the number of observations the object receives. Using additional number fitted values in other bands would not add any additional benefit and will contribute to noise to the feature set. Therefore they have been excluded when training the autoencoder.} and footnote \footnote{The Flags value has been input as an integer. Given the limited number of flags present in the data, it was not necessary to introduce further encoding of this feature.}
\end{deluxetable}

 Exploring the full extent of the feature space is an ideal area for future work. At this point most of the features can be considered static and neglect to a large extent, the dynamical changes of an object. As LSST will observe the 5 million small body discoveries hundreds of times \citep{2009arXiv0912.0201L,2019ApJ...873..111I}, there will be a rich set of time series data to utilise.
 Nonetheless, analysis of measurements over time may still uncover extremely interesting objects.

A data pre-processing pipeline was created before training the encoder-decoder network. Firstly, objects with missing features were dropped. This accounted for 30$\%$ of objects. The majority of these objects were dropped because they simply do not have enough observations in specific filters to produce a reliable fit and derive magnitudes. We do not consider imputation of missing features in this paper. However, dealing with missing data is an important problem to address in future work. It is especially salient for enabling object anomaly detection during the course of survey operations as some features may not be available given a limited number of detections.

The data were normalised to the range $0 \le X'_i \le 1$, where $X'_i$ is a single input feature vector, in order to allow the sigmoid activation to be used in final layer of the autoencoder. The data were subsequently split into training (80$\%$), validation (10$\%$), and test sets (10$\%$).

Using the Keras package \citep{chollet2015keras, 10.5555/3203489}, a deep autoencoder was constructed. A number of architecture decisions were made regarding the autoencoder's encoder and decoder networks. The encoder and decoder were chosen to be symmetric, each with three hidden, dense layers. Each layer used the ReLU activation function apart from the final layer of the decoder, which applied a sigmoid activation function to ensure the output was normalized. Batch normalisation was also applied after each layer's activation function.

The autoencoder was trained using the Adam optimizer \citep{kingma2017adam}. The loss function consisted simply of minimising the mean square error between the input batch and the output of the encoder, $e$ and decoder, $d$. The difference between a single instance's input and output from an autoencoder is referred to as the reconstruction loss. The loss function is the mean reconstruction loss over the batch, which is written as 

\begin{equation}
    loss = \frac{1}{N} \sum_{i=1}^{N}(X_i - d(e(X_i))^2
\end{equation}

where N is the batch size used in single training epoch. The validation loss was monitored using a callback function to prevent the network overfitting to the training set.

The units of the autoencoder's layers and the learning rate were determined with Bayesian optimisation using Keras Tuner \citep{omalley2019kerastuner}. This was performed for latent space dimensions of size 2, 4, 6, 8 and 10. For each latent space dimension, the network was trained over five optimisation trials using different combinations of units. The best hyper-parameter configurations from the Bayesian optimization step were stored and used to retrain the networks. Table \ref{table:latent_tradeoff} shows how each latent space component performs on the reconstruction loss and its compression ratio on the test set. Here the compression ratio is the size of the latent representation compared to the original input data.

\begin{deluxetable}{|l|c|c|c|}
\label{table:latent_tradeoff}
\tablehead{
\colhead{Latent dimensions} & \colhead{Architecture} & \colhead{Mean reconstruction loss} & \colhead{Compression ratio}
}
\startdata
\hline
 2 & 108-52-29-2-29-52-108 &$3.35 \times 10^{-4}$ & 0.029 \\
 \hline
 4 & 125-59-16-4-16-59-125 &$3.65 \times 10^{-5}$ & 0.057 \\
 \hline
 6 & 97-34-24-6-24-34-97 &$7.73 \times 10^{-6}$ & 0.086 \\
 \hline
 8 & 88-56-27-8-27-56-88 &$6.32 \times 10^{-6}$ & 0.114  \\
 \hline
 10 & 90-60-32-10-32-60-90 &$5.42 \times 10^{-6}$ & 0.143 \\
 \hline
\enddata
\caption{The optimized architecture for each latent space dimension, presented with the resulting reconstruction loss performance on the test set. The compression ratio measures the size of the latent representation compared to the original dataset.}
\end{deluxetable}

Six latent dimensions were selected because this implementation had the most optimal reconstruction loss and compression ratio. Performance improvements degraded after including more components which indicates that additional components contributed noise to the latent representation. This suggests that six components is the optimal representation size for this feature space. However it could also indicate deficiencies in the training stages of the eight and ten latent component networks. Given GPU constraints, the limit of five trials means that these networks may not have found the most optimal unit configuration. With greater resources this problem could be resolved using more trials of the optimization procedure. However we will adopt six latent components for the course of this analysis.

Despite the limitations of the larger latent space representations, the best model was then compared to two other dimensionality reduction techniques: principal component analysis (PCA) and non-negative matrix factorisation (NMF) \citep{NMF}. With a reconstruction loss of $7.73 \times 10^{-6}$ on the test set, the autoencoder outperformed PCA by $81\%$ and NMF by $92 \%$. The improvement over these benchmarks provides confidence that the autoencoder can be used as an effective dimensionality reduction method on the object feature space.

\begin{figure}
    \label{fig:ae_architecture}
    \centering
    \includegraphics[width=\textwidth]{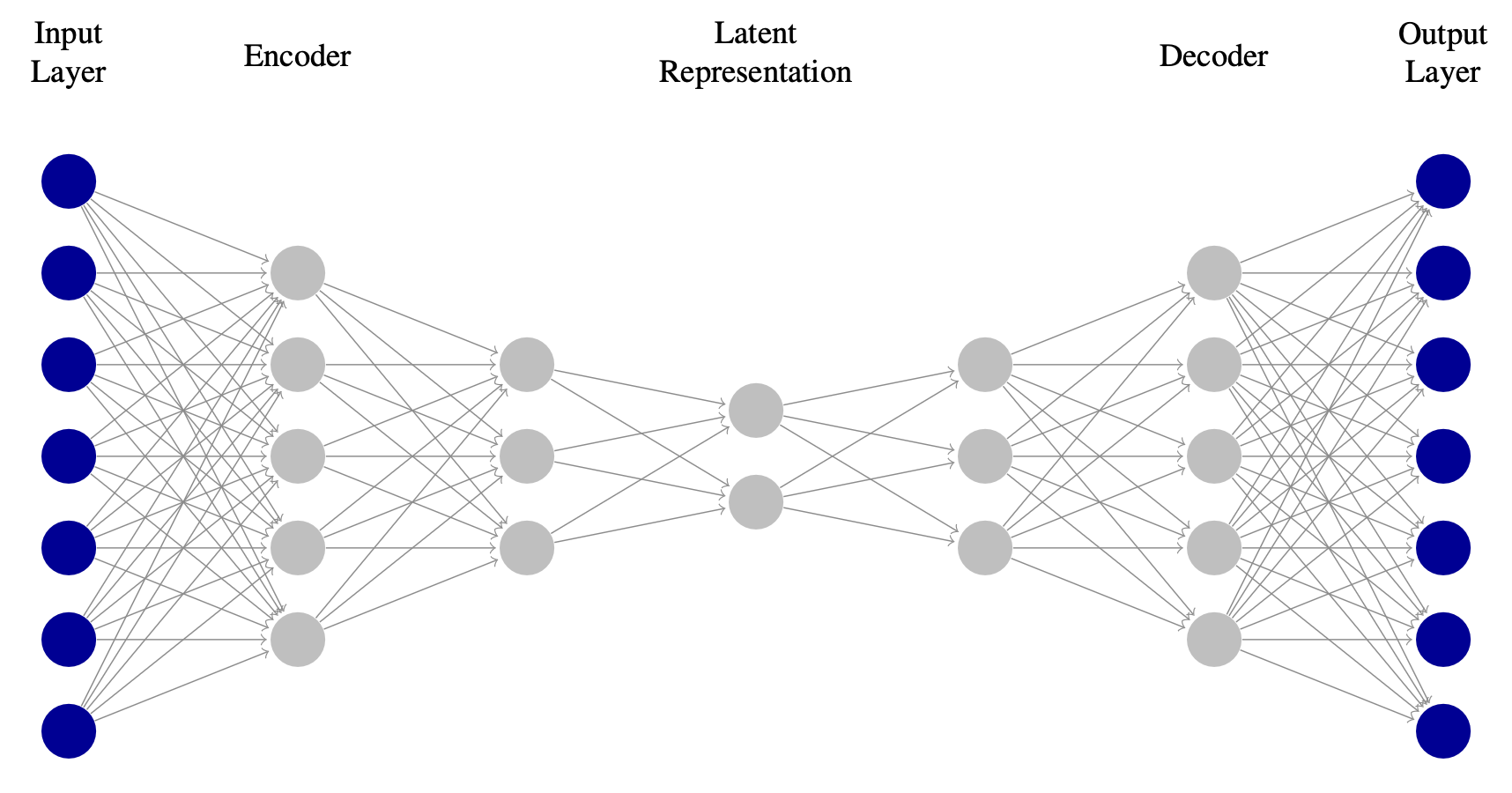}
    \caption{A schematic diagram of the general network architecture of an autoencoder. The input layer is followed by the encoder which is made made up of a number of hidden layers. In the centre lies the latent representation which can be transformed into an output using the decoder.}
    \label{fig:ae_bench}
\end{figure}

\subsection{Reconstruction loss}
\label{section:recon_loss}
A fundamental feature of deep autoencoders that is useful for anomaly detection tasks is the reconstruction loss. 
By encoding and decoding an object's feature vector, we can measure the error between the input and output from the network. Normal samples are reconstructed well by the encoder-decoder network and hence have low reconstruction losses. More outlying examples have higher reconstruction losses, and therefore this loss can be used as an anomaly score.

\begin{figure}
\gridline{\fig{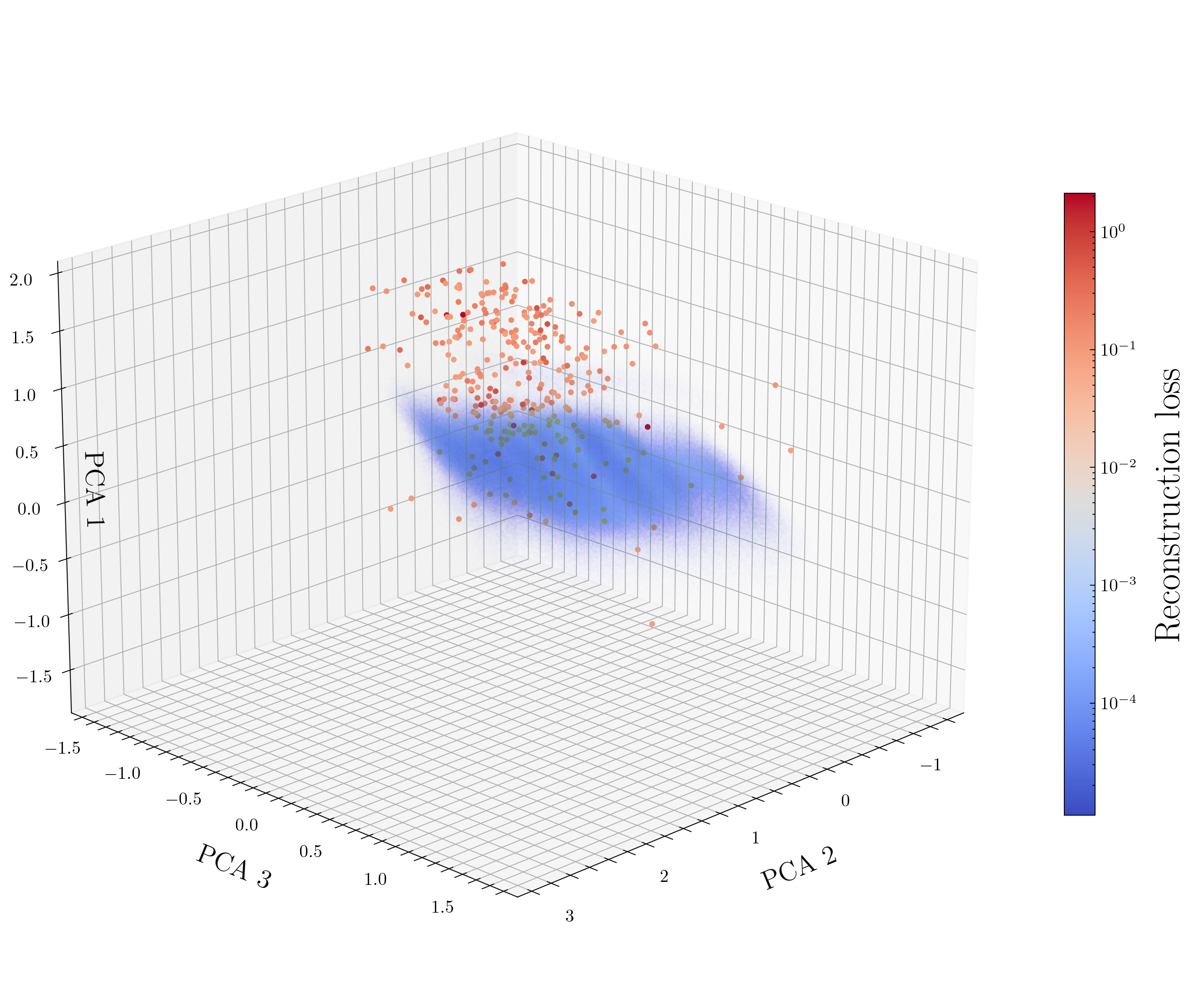}{\textwidth}{(a)}}
\gridline{\fig{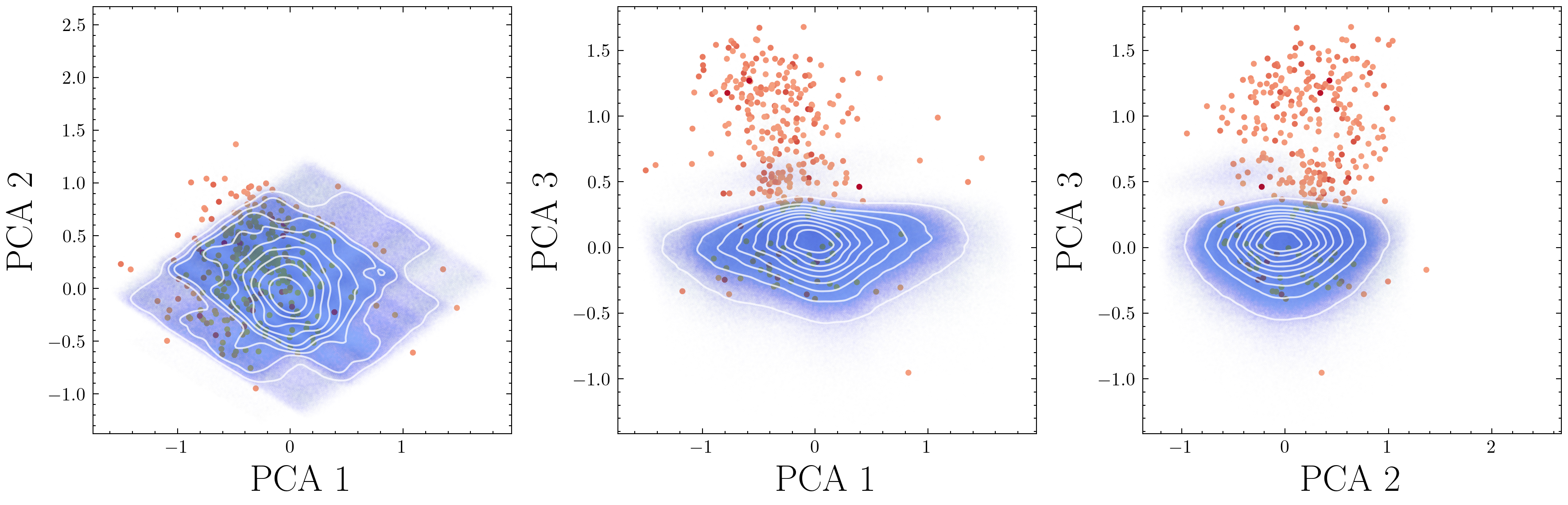}{\textwidth}{(b)}
          }
\caption{Reconstruction scores for 3.1 million Solar System objects across the reduced feature space. The colour bar describes the reconstruction loss. More anomalous points are deeper in the red. The top 0.01$\%$ anomalies are enlarged for this plot. They lie distant from the majority of normal objects in blue.}
\label{fig:recon_loss_feature_space}
\end{figure}

Using PCA, the latent components were reduced to three dimensions which could be plotted. Figure \ref{fig:recon_loss_feature_space} demonstrates the reconstruction loss values across the feature space, which are redder for higher reconstruction losses. Anomalous objects tend to lie in less dense regions of the reduced feature space, away from the points in blue that are reconstructed well by the autoencoder. Global anomalies become apparent as they are distant from the low reconstruction blue instances. In some regions we see that some instances group together in clusters. We also see the fingerprints of local anomalies appearing towards the edges of more dense regions. This highlights that the anomaly prototypes defined in section \ref{section:GMM} are not exclusive to the orbit-colour space, but rather that they are an abstract way of defining anomalies in a general feature space.

As these anomalous instances separate from normal points in the latent space, we can explore these regions to find potentially interesting objects. This will be the focus of our next section.

\subsection{Similarity searching}
\label{section:sim_search}
We saw directly in Section~\ref{section:recon_loss} that objects with higher anomaly scores reside close to each other in the latent space, usually very distant from normal objects. Pairing this with the inherent feature of the autoencoder's latent space, that objects with similar properties will  lie close to one another, means we can look at the nearest neighbors of anomalous points to find more examples. Further, we can also use similarity searches to help explain why such an object is unusual. On one hand, this helps to address the problem of the interpretabilty of the latent space. More importantly on the other, we can build helpful search and recommender systems to assist in finding the most unusual objects discovered by LSST.

\begin{figure}
    \centering
    \includegraphics[width=0.9\textwidth]{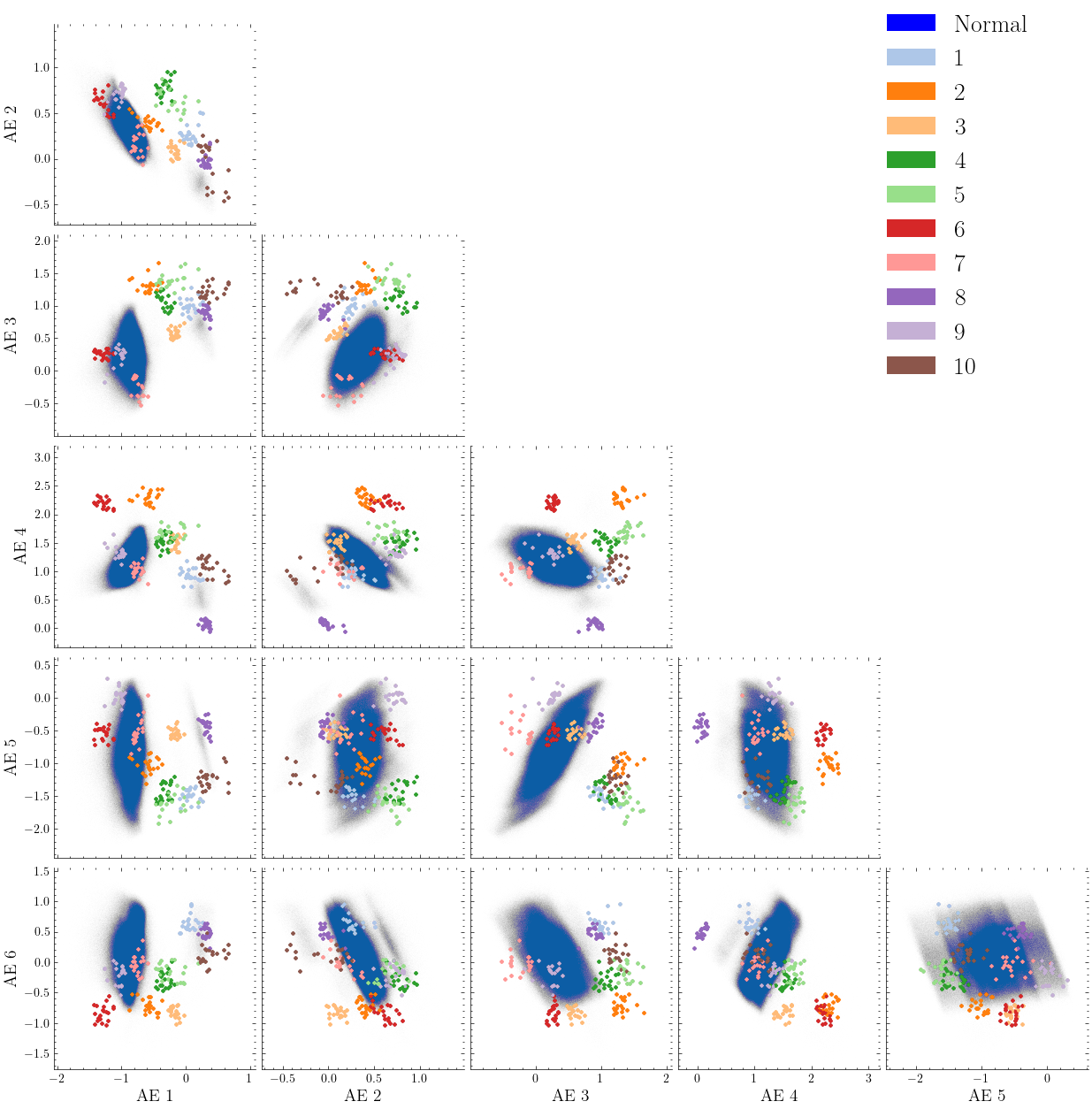}
    \caption{The latent space representation of the top ten anomalies and their nearest neighbors. These neighborhoods of anomalies are colour coded and lie outside the dense blue distributions of normal objects in the latent space.}
    \label{fig:anom_space}
\end{figure}

To illustrate the usefulness of similarity searching, we found the top ten anomalies ranked by reconstruction loss across 3.1 million Solar System objects in the ten year survey simulation. After locating their position in the latent space, we found each object's twenty nearest neighbors. We plotted the neighborhood of points associated with these ten anomalies in figure \ref{fig:anom_space}; each of the color-coded clusters can be clearly seen to occupy particular regions of the plots. In this figure each axis represents a latent space component; for example `AE 1' is the first latent space component of the autoencoder.

Each neighborhood of objects has specific anomaly characteristics. Neighborhoods 1-5 and 10 show extreme filter magnitudes and respective uncertainties in their measured magnitudes. Many of these magnitudes are beyond the detection limit of the survey and represent unusual artifacts of the survey simulation. These objects are set to be fixed for future simulations. This highlights that this approach can help to find bugs in future survey simulations. 

Neighborhood 6 was completely comprised of ISOs. ISOs are rare objects in the simulation. This subset is particularly unusual because of their rare eccentricities. The most anomalous of these ISOs had $e = 10.78$. This object was confirmed to have the highest eccentricity in the processed data. The average eccentricity of the ISOs in this neighborhood was e = 6.50 $\pm$ 1.72 with perihelion of q = $2.03 \pm 0.44$ au.

Neighborhood 7 was a collection of objects with highly inclined and eccentric orbits. At the centre of this neighborhood was a trans-Neptunian object with inclination of $i = 169.1 ^\circ$, $e = 0.95$ and $a = 35.9$ au. The mean inclination of the neighborhood was $i = 84.85 ^\circ$ whilst the mean eccentricity was $e = 0.79$. This further demonstrates that objects with high eccentricities, albeit not as outlying in terms of their eccentricities as the interstellar objects found in neighbourhood 6, are also grouped close in the latent space. With the high eccentricity of this group, there were found to be six Mercury crossers including the real Solar System objects (277142) 2005 $LG_8$, (494706) 2005 $GL_9$, (136874) 1998 $FH_{74}$, and (24443) 2000 $OG$. The Mars crossing asteroid (4257) Ubasti was also in this neighborhood because of its highly eccentric and inclined orbit. The Earth crosser (3752) Camillo which is an outer grazer with a highly inclined orbit was also a neighbor. Unlike neighbourhood 6, this seems a more counter intuitive mix of objects. However given the orbital characteristics, in particular the eccentricities of the objects, this is a logical group of objects which a human may miss.

Neighborhood 8 consisted of outlier objects in the vicinity of Jupiter. The most anomalous of this neighborhood were comets, LPCC1411 and LPCD1330. The latter, in particular, belongs to the rare class of long period comets in DP0.3. Both objects had extreme inclinations of $i = 160.8 ^\circ$ and $i = 119.0 ^\circ$, with $q = 4.40$ au and $q = 5.84$ au. The rest of the neighborhood consisted of 18 asteroids with $a = 5.20 \pm 0.04$ au, making many of these objects Jupiter trojans. They have higher than average $i - z$ colours and inclinations compared to the wider trojan population. 

Neighborhood 9 consisted completely of non-periodic (C) type comets that were highly inclined to the ecliptic. The average inclination of this neighborhood was $i = 149.2 ^\circ$.

The summaries of these neighborhoods presented here and in figure \ref{fig:anom_space} demonstrate the variety of interesting objects we can now detect and discover using the autoencoder. We now move to consider how we can narrow the search for specific anomaly types using a model based outlier method in conjunction with the autoencoder.

\subsection{Improved anomaly search using an ensembling approach.}
\label{section:ae_ensembling}
We will consider how the reconstruction loss can be used in conjunction with another model, narrowing the anomaly type returned. By considering multiple models, we can improve on any weakness in any one particular model. This can be considered a simple example of model ensembling. We return to the GMM model of the orbit-colour space developed in section \ref{section:GMM}. To demonstrate this hybrid approach, we consider a region defined by a semi-major axis range spanning 5.16 au to  5.24 au. This region has approximately 40,000 objects, primarily consisting of Jupiter trojan asteroids.

\label{section:weird_orbit_colour}
\begin{figure}
    \centering
    \includegraphics[width=\textwidth]{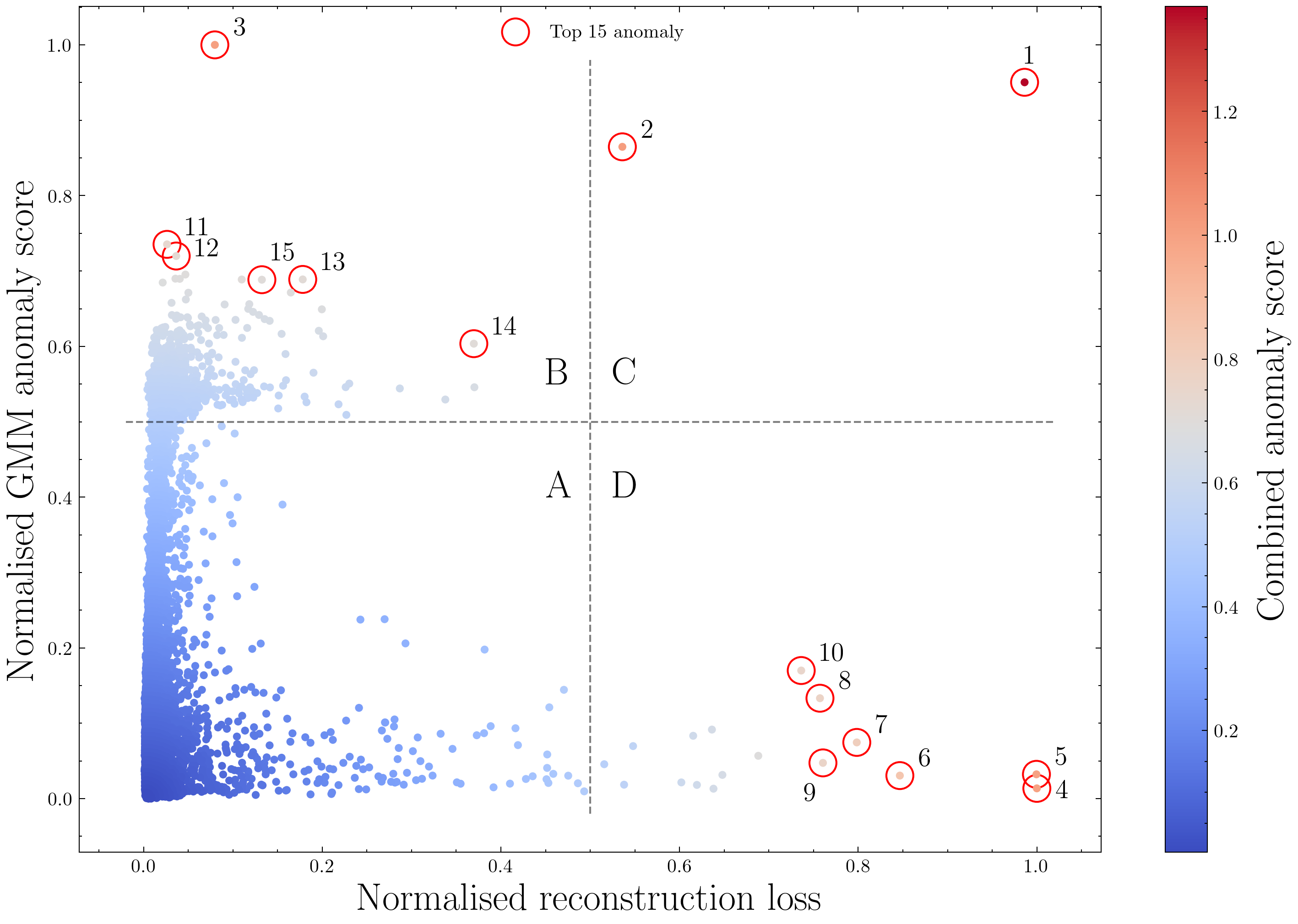}
    \caption{Visualisation of the normalised autoencoder reconstruction loss versus normalised anomaly score from the GMM. It can be seen with points (4), (5) that outlying objects are suggested that would be missed by considering only the model based approach.
}
    \label{fig:oc_anom_space}
\end{figure}

One particular issue with domain specific anomaly detection is that it neglects the wider context of the object. An object may not be so anomalous when only considering the five parameters of the orbit-colour space, yet it may be much more interesting when we consider other features. To mitigate this, we combine the GMM anomaly detection model with the reconstruction loss. Using the GMM, we can find the most unlikely objects in the orbit-colour space using the log-likelihood. By computing the Euclidean distance in this 2D anomaly space, we calculate a combined anomaly score. This allows recovery of objects that are anomalous in terms of both their orbit-colour properties and in their wider context across the entire feature space.

Broadly there are 4 regions of figure \ref{fig:oc_anom_space}, shown by the quadrants placed at the center of each axis.  The bottom left quadrant (A) has normal objects that are not scored highly by either method, containing over $97\%$ of all objects. These are the bulk of of objects that can be safely ignored for anomaly searches.

The top right quadrant (C) is objects rated highly by both methods. These are priority objects for review accounting for less than $0.01\%$ of all objects. Object (1) has the lowest $a^*$ value of any asteroid considered in this region. With a high $i - z$ colour and low eccentricity the GMM model rates this as a high priority object. Further it has high reconstruction loss compared to the trojan population having the largest but highly uncertain $G_{12}$ slope parameter in the $r$ band. Object (2) conversely has the highest $a^*$ colour amongst the trojans with an inclined orbit rating highly in outliers in the orbit-colour space. With largest $G_{12}$ slope parameter error in the $g$ band it too has large uncertainties in this band, which is reflected in the high reconstruction loss. Both objects have some of the smallest absolute magnitudes in the population.

The upper left region (B) contains objects that rank low in reconstruction loss but higher in the GMM anomaly score. These have properties we expect from orbit-colour outliers and account for $2.86\%$ of all objects. Objects (11) and (12) have high inclinations and outlying $i - z $ colours. Objects (14) and (15) have extreme $i - z$ colours whilst (13) has a combination of extreme colours in both $a^*$ and $i - z$. Given their low reconstruction losses, these objects will tend to be normal in other features. For users interested in narrow anomaly searches this region should be searched thoroughly as these objects will tend to be anomalous according to the specifics of the model.

The bottom right region (D) has objects low in model score but high in reconstruction loss. These are objects outlying in other features except their orbit-colour properties. It contains $0.046\%$ of all objects. What is especially notable is that these asteroids have a mean MPC magnitude of $\bar{H} = 9.35$ which is $5 \sigma$ from the rest of the trojan population.

Each region of this graph can be used for different purposes when we consider more general ensembling approaches. Quadrant (C) can be used to find anomalous instances that are high priority for review. Detection models can be improved by looking at quadrants (B) and (D) for where there is disagreement in models, which can yield interesting results or reveal blind spots in models. In this case the magnitude was neglected by the GMM model but recovered by the autoencoder model.

\section{Conclusion}
This paper has explored various methods for anomaly detection for LSST Solar System object data. Given the deep autoencoder's breadth of application, we believe this is a promising approach going forward. By providing anomaly scores with the reconstruction loss it can be used directly for unsupervised anomaly detection. It can also be combined with other finer methods to discover more specific anomaly types. Additionally, the autoencoder's compressed but meaningful latent representation of the features of Solar System objects is searchable and interpretable. By building efficient user interfaces on top of this search mechanism, we can easily find similar objects, anomalous or normal for users to examine. In this way the Solar System catalog becomes much more navigable, even as it expands to unprecedented sizes during the lifetime of LSST.

As well as identifying natural anomalous objects, this work could be applied to technosignature searches within LSST Solar System data \citep{1980Icar...42..442F, 2021ApJS..257...42L}. Any search for technosignatures may also uncover objects which initially appear artificial, but which upon further study turn out to have natural astrophysical explanations.

Beyond looking for outliers in orbit-colour space, anomaly searches can leverage additional dimensions of the LSST data set. \citet{2019PASP..131h4401L} considers how a technosignature may be detected using prospective surface `glints', likely manifesting as a single or series of outlier measurements in the phase and light curves of the object. This is an opportune application for extending the anomaly detection approaches suggested in this paper using LSST data products, and has direct synergies with the search for outbursts of activity in small Solar System bodies, along with other interesting phenomena. Currently light curves have not been fully simulated in the data products used in this paper; however, they are planned to be included in future simulations. 

After evaluating the deficiencies of standalone unsupervised methods, we demonstrated the power of human feedback in detecting anomalies in section \ref{section:deficiencies_unsupervised} by using a supervised approach. Using human feedback can increase relevance, accuracy and precision of the anomaly detection system. An approach like Astronomaly \citep{Lochner_2021} provides inspiration for application to LSST data. Intuitive user interfaces that display interesting objects detected by unsupervised methods can be rated by users. These ratings can be used to train supervised algorithms which can be personalised to anomaly type or to a user preference. By putting the right anomalies in the right hands we can multiply the value of the data collected by LSST and precipitate potential follow up studies for the most interesting objects found in the survey. We have demonstrated that deep autoencoders can fulfil this role as an unsupervised detection model by performing on the scale of LSST and that they can enable efficient anomaly discovery for the most interesting Solar System objects. 
\section{Software and third party data repository citations}
\software{
DiRAC LSST Solar System platform,
Rubin Science Platform,
Matplotlib \citep{Hunter:2007},
Seaborn \citep{Waskom2021},
NumPy \citep{harris2020array},
Pandas \citep{reback2020pandas},
Sci-Kit learn \citep{scikit-learn},
PyOD \citep{zhao2019pyod}.
}\\
The Solar System Products database ({\url{https://lsst.dirac.dev/}})\\
The DP0.3 simulation (\url{https://dp0-3.lsst.io})\\

\begin{acknowledgments}
BR and SC were supported by Breakthrough Listen, which is managed by the Breakthrough Initiatives, sponsored by the Breakthrough Prize Foundation\footnote{\url{http://www.breakthroughinitiatives.org}}. MES was supported by the UK Science Technology Facilities Council (STFC) grant ST/X001253/1. CJL acknowledges support from the Alfred P. Sloan foundation.

JRAD acknowledges support from the DiRAC Institute in the Department of Astronomy at the University of Washington. The DiRAC Institute is supported through generous gifts from the Charles and Lisa Simonyi Fund for Arts and Sciences, and the Washington Research Foundation.

This research has made use of NASA's Astrophysics Data System Bibliographic Services.
The authors thank the Rubin Solar System Processing, Database, the Science Platform, and Community Science Teams for their support of Rubin Observatory Data Preview 0.3. This material or work is supported in part by the National Science Foundation through Cooperative Agreement AST-1258333 and Cooperative Support Agreement AST1836783 managed by the Association of Universities for Research in Astronomy (AURA), and the Department of Energy under Contract No. DE-AC02-76SF00515 with the SLAC National Accelerator Laboratory managed by Stanford University.    

This work made also made use of the resources provided by LINCC Jupyter Hub (https://lsst.dirac.dev). The LINCC Jupyter Hub is powered by Microsoft Azure, made available through the UW eScience Institute's Cloud Credits program. LINCC Frameworks is supported by Schmidt Futures, a philanthropic initiative founded by Eric and Wendy Schmidt, as part of the Virtual Institute of Astrophysics (VIA).

Data Access:  The contents of Rubin Observatory's Data Preview 0.3 used in this paper are available from the Vera C. Rubin Observatory Construction Project and Operations Teams via the Rubin Science Platform \url{https://data.lsst.cloud/}.

\end{acknowledgments}
\vspace{5mm}
\bibliography{main}{}
\bibliographystyle{aasjournal}
\end{document}